\documentclass[aps,prl,reprint,superscriptaddress]{revtex4-2}

\usepackage[english]{babel}
\usepackage[utf8]{inputenc}
\usepackage[top=2cm, bottom=2cm, left=2cm, right=2cm]{geometry}
\usepackage{color,soul}
\usepackage{graphicx}

\usepackage[svgnames]{xcolor}
\usepackage{amsmath}
\usepackage{tabulary}

\begin{document}

\title{Mechanism of Indirect Photon-induced Desorption at the Water Ice Surface}

\author{R. Dupuy}
\email{remi.dupuy@obspm.fr}
\author{M. Bertin}
\author{G. F\'{e}raud}
\author{X. Michaut}
\author{P. Marie-Jeanne}
\author{P. Jeseck}
\author{L. Philippe}
\affiliation{Sorbonne Universit\'e, Observatoire de Paris, Universit\'e PSL, CNRS, LERMA, F-75005 Paris, France}
\author{V. Baglin}
\affiliation{CERN, CH-1211 Geneva 23, Switzerland}
\author{R. Cimino}
\affiliation{Laboratori Nazionali di Frascati (LNF)-INFN I-00044 Frascati}
\author{C. Romanzin}
\affiliation{Laboratoire de Chimie Physique, CNRS, Universit\'e Paris-Sud, Universit\'e Paris-Saclay, 91405, Orsay, France}
\author{J.-H. Fillion}
\affiliation{Sorbonne Universit\'e, Observatoire de Paris, Universit\'e PSL, CNRS, LERMA, F-75005 Paris, France}

\begin{abstract}

Electronic excitations near the surface of water ice lead to the desorption of adsorbed molecules, through a so far debated mechanism. A systematic study of photon-induced indirect desorption, revealed by the spectral dependence of the desorption (7-13 eV), is conducted for Ar, Kr, N$_2$ and CO adsorbed on H$_2$O or D$_2$O amorphous ices. The mass and isotopic dependence and the increase of intrinsic desorption efficiency with photon energy all point to a mechanism of desorption induced by collisions between adsorbates and energetic H/D atoms, produced by photodissociation of water. This constitutes a direct and unambiguous experimental demonstration of the mechanism of indirect desorption of weakly adsorbed species on water ice, and sheds new light on the possibility of this mechanism in other systems. It also has implications for the description of photon-induced desorption in astrochemical models.

\end{abstract}

\maketitle


Water ice is ubiquitous on Earth and in space, and often plays the role, among others, of an environment onto which less abundant molecules adsorb, desorb and react \cite{vandishoeck2013,bartels-rausch2012,hama2013,kolb2010,george2015}. Energetic irradiation of such environments will result in energy deposition mostly in the form of electronic excitations of the water matrix, which can subsequently affect molecules adsorbed on the ice through energy transfer or reaction with radicals. Adsorbate desorption is one of the possible outcomes. Photodesorption by vacuum-UV (VUV) photons is, for instance, an important process taking place at the surface of interstellar dust grains, which are covered by an icy mantle mainly composed of water, along with smaller amounts of molecules such as CO, CO$_2$, N$_2$ or CH$_4$ \cite{boogert2015}. It affects the gas and ice molecular abundances in the interstellar medium, and explains how some molecules (such as water) are observed in the gas at temperatures where their thermal desorption from grains is not possible \cite{hogerheijde2011,caselli2012}. 

A molecular-scale description of the mechanism by which an electronic excitation eventually causes desorption has so far been achieved only for a handful of molecular or atomic solids, such as rare-gas solids \cite{runne1995,reimann1988}. This is despite a considerable amount of experimental measurements on photodesorption focusing on model "pure" molecular ices \cite{munozcaro2018a,mccoustra2017,oberg2016,yabushita2013}, including water ice, as detailed below. On the other hand, there is currently a lack of detailed understanding of the desorption properties - both the desorption mechanism and an estimate of the yield of the process - of more realistic "mixed" systems. In such systems, indirect desorption processes - where an excited matrix species transfers energy to a different species which desorbs - play an important role. Indirect photodesorption has been studied for the astrophysically relevant CO-ice environment \cite{bertin2012,bertin2013,dupuy2017a,feraud2019a,carrascosa2019} but the exact mechanism remains elusive. Water ice is a more complex but very important system for which indirect desorption has been little explored. 
 
What has been the object of many studies is the fate of electronic excitations at the pure water ice surface \cite{johnson2011,ramaker1983a,yabushita2013,dartois2015,marchione2016a,yabushita2006,miyazaki2020}, and how they may lead to desorption of intact water \cite{andersson2008,arasa2015,yabushita2009,nishi1984,desimone2013, fillion2020,westley1995,cruz-diaz2018,oberg2009c,arasa2010,andersson2011}. The dominant underlying mechanism for this latter process remains nonetheless debated. Theoretical works \cite{andersson2008,arasa2015} suggested the co-existence of two different mechanisms, initiated by water photodissociation: one where an energetic H atom issued from the dissociation transfers momentum to another molecule and "kicks" it out into vacuum, and another where a dissociated water molecule immediately recombines (geminate recombination) with the exothermic energy of recombination leading to desorption. Experimental measures of the rotational and kinetic energies of desorbing H$_2$O (v=0) were shown to be compatible with these mechanisms \cite{yabushita2009}. However, other experimental and theoretical investigations \cite{nishi1984,desimone2013}, in principle very similar, favoured instead a more direct mechanism, without dissociation but involving exciton localization at the surface and an electronic repulsion of the excited water molecule relative to its surroundings, causing its desorption, in line with the classic MGR (Menzel-Gomer-Redhead) model \cite{avouris1989}. In addition, we showed in a recent study \cite{fillion2020} that the observed isotopic and temperature effects in the desorption from water ice required consideration of additional mechanisms, such as desorption channels linked with photochemistry. From the available experimental evidence, it is therefore not possible to conclude definitively on the existence of any of these mechanisms and whether they can be extrapolated to other closely related systems. The desorption of other molecules adsorbed on water ice, for instance, remains largely unexplored, despite its importance for a more realistic description of electronic desorption on actual icy surfaces.

In this letter we report a study of indirect VUV photon-induced desorption of molecules weakly adsorbed (E$_b$ $<$ 130 meV) on amorphous water. We systematically measured indirect photodesorption yields for four adsorbates (Ar, Kr, CO, N$_2$) on either H$_2$O or D$_2$O ices. Our results reveal unambiguously that the energetic H atom collision mechanism dominates indirect desorption for the systems investigated here. It is therefore now clear that there are systems where this mechanism does operate and even dominates, although it is not the case for water desorption from the pure ice \cite{fillion2020,desimone2013}.

The experiments were performed in an ultra-high vacuum chamber coupled to the DESIRS undulator beamline \cite{nahon2012} of the SOLEIL synchrotron which provides monochromatic photons in the VUV range (7-13 eV). Ices are grown on a polycrystalline copper substrate cooled down to 15-100 K by exposition to a partial pressure of gas using a dosing tube. First a 100 ML$_{eq}$ (monolayer equivalent) water ice is grown at 100 K, yielding compact amorphous solid water \cite{stevenson1999} (c-ASW). 1 ML$_{eq}$ of the adsorbate is then deposited on top at a temperature of 15 K. The thickness of the deposited molecular ice is controlled and calibrated using the temperature programmed desorption (TPD) technique \cite{doronin2015}. Photodesorption induced by irradiation of the ice is monitored in the gas phase by quadrupole mass spectrometry (QMS). We have only monitored desorption of the adsorbates, as intact water, OH or H fragments cannot be detected under these experimental conditions. The QMS is calibrated to derive absolute photodesorption yields, in molecules/incident photon. More experimental details are described in previous papers \cite{dupuy2017a,dupuy2018c} and in the Supplemental Material, which includes ref. \cite{doronin2015a,fayolle2011,orient1987,rejoub2002,martin-domenech2015,cruz-diaz2014}.   

\begin{figure}
    \includegraphics[trim={0cm 0cm 0cm 0cm},clip,width=\linewidth]{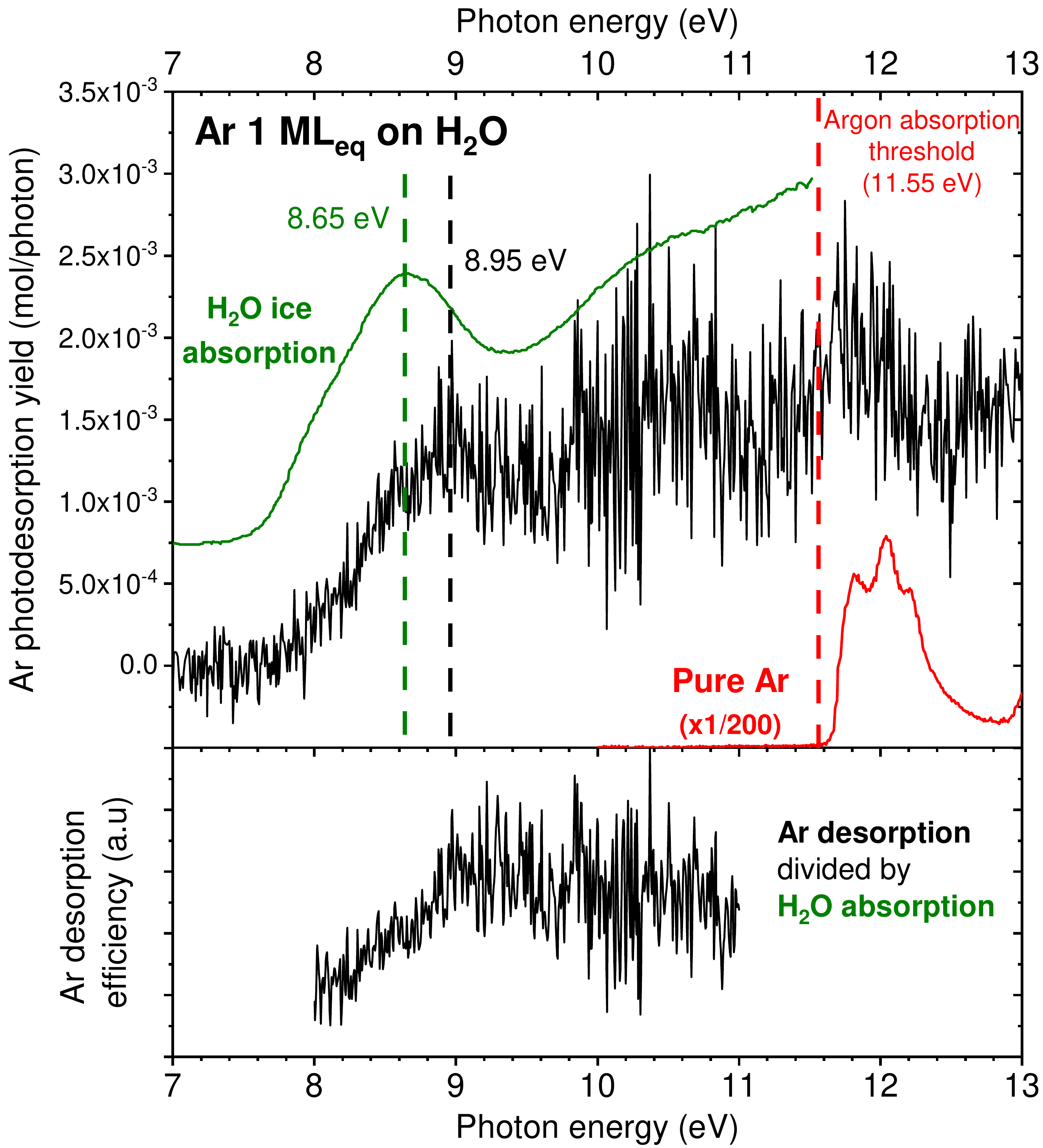}
    \caption{Upper panel: photodesorption spectrum (desorption yield as a function of photon energy) of argon for 1 ML$_{eq}$ of Argon adsorbed on 100 ML compact amorphous solid water c-ASW (black line). Also shown for comparison are the absorption spectrum of amorphous H$_2$O ice from Lu et al. \cite{lu2008} (green line, arbitrary units) and the photodesorption spectrum of Argon from 20 ML pure Argon ice (red line, scaled 1/200). Lower panel: photodesorption spectrum of argon 1 ML$_{eq}$ on c-ASW (black spectrum of the upper panel) divided by H$_2$O absorption (green spectrum of the upper panel).}
    \label{Ar_H2O}
\end{figure}      

The photodesorption spectrum of Ar from the Ar/H$_2$O system is presented on fig. \ref{Ar_H2O}. Also shown for comparison are the photodesorption spectrum of Ar for a pure 20 ML Ar ice, and the absorption spectrum of amorphous water ice taken from Lu et al. \cite{lu2008}. Spectral information provides here a direct evidence of indirect desorption. The pure Ar photodesorption spectrum is very different from the spectrum of Ar desorbing from Ar/H$_2$O: the threshold for Ar absorption, and desorption from the pure ice, is 11.55 eV. Desorption of Ar below that photon energy in the Ar/H$_2$O spectrum is therefore due to water ice excitation, as is seen from the resemblance between that part of the desorption spectrum and the absorption spectrum of water ice. We will focus here in particular on the first electronic excited state of condensed water, which is dissociative in the gas phase (where it corresponds to the 4a$_1$ ← 1b$_1$ transition) and in condensed phase \cite{yabushita2006}. An apparent shift of the maximum of this state, observed at 8.65 eV in water ice absorption and at 8.95 eV in the Ar from Ar/H$_2$O desorption spectrum, can be observed. 

\begin{figure}
    \includegraphics[trim={0cm 0cm 0cm 0cm},clip,width=\linewidth]{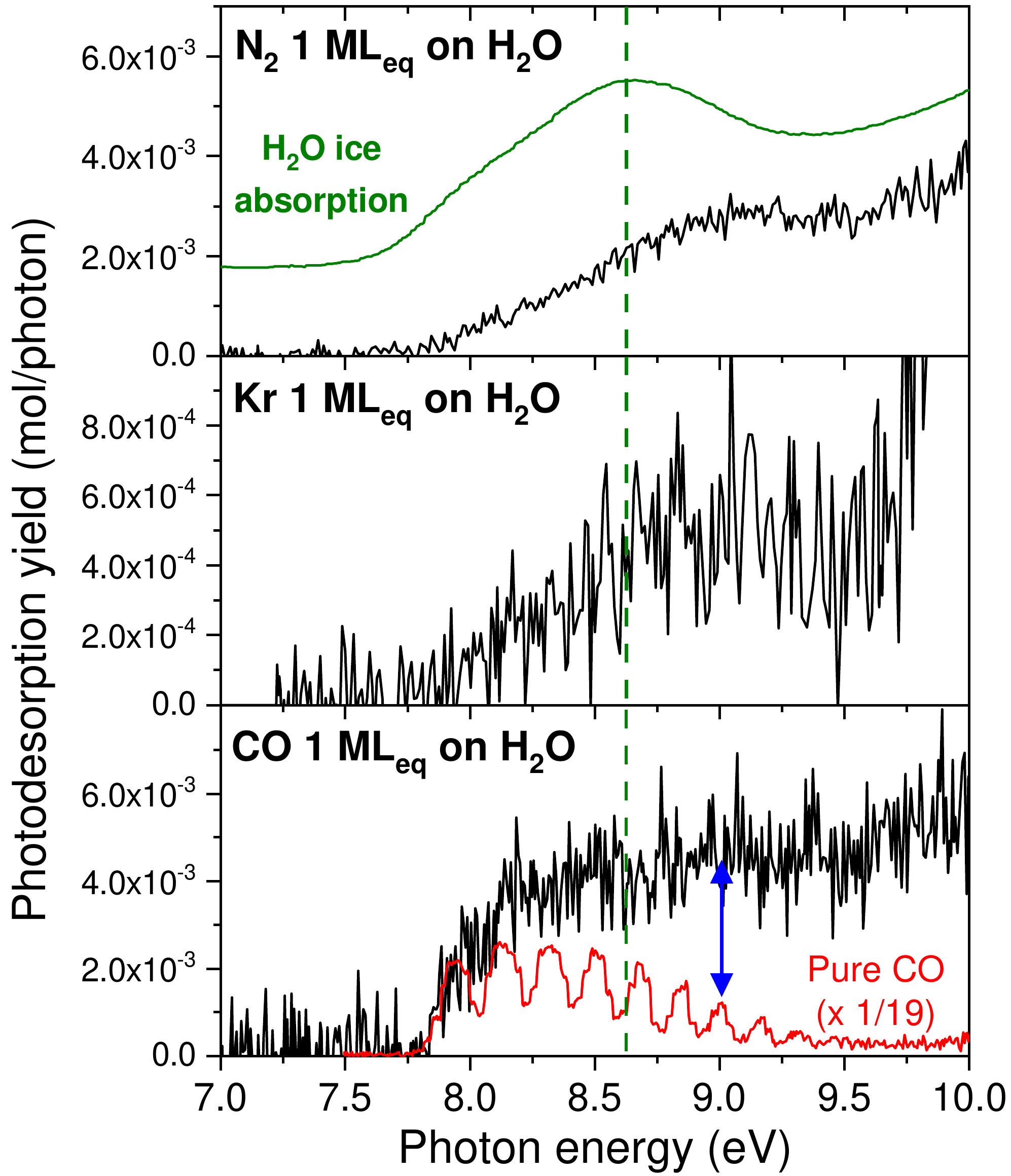}
    \caption{Photodesorption spectra (black lines) of $^{15}$N$_2$ from $^{15}$N$_2$ adsorbed on c-ASW (top panel), Kr from Kr adsorbed on c-ASW (middle panel) and $^{13}$CO from $^{13}$CO adsorbed on c-ASW (bottom panel). The absorption spectrum of c-ASW from Lu et al. (green line) is reproduced on the top panel. On the bottom panel, the photodesorption spectrum of 20 ML pure CO is also shown (scaled 1/19) and fitted under the CO/H$_2$O photodesorption spectra to estimate the contribution of indirect desorption, indicated by the blue arrow.} 
    \label{layers_H2O}
\end{figure}

This apparent shift actually corresponds to a deformation of the spectrum caused by an increase of desorption efficiency throughout the first electronic state of H$_2$O ice. This is visible in the lower panel of fig. \ref{Ar_H2O} where the Ar photodesorption spectrum was divided by the H$_2$O absorption spectrum, yielding the intrinsic desorption efficiency. The efficiency increases almost linearly from 8 to 9 eV, then remains constant at higher photon energy.

Similar experiments performed on 1 ML$_{eq}$ Kr, N$_2$ and CO adsorbed on compact amorphous H$_2$O are displayed in fig. \ref{layers_H2O}. The absolute photodesorption yields differ but the spectral shape for Kr and N$_2$, whose first dipole-allowed electronic state are respectively at 9.9 eV and 12.4 eV in the condensed phase, is similar to Ar/H$_2$O. For both these adsorbates, as for Ar, it is therefore straightforward to measure indirect desorption through the spectral signature of the first water ice excited state around 9 eV. This state appears again blue-shifted compared to the absorption spectrum in the desorption spectra. The excitation threshold of CO on the other hand is at 7.9 eV, and its first excited state overlaps with the water ice excited state. Still, the structure due to the vibronic bands of that CO excited state makes it so that one can still easily see the presence, and estimate the quantitative contribution, of indirect water-induced desorption to the overall CO desorption spectrum.

The experiments were repeated for the same adsorbates on D$_2$O ice (grown under the same conditions) for comparison. The results are shown in fig. S2 and are quite similar to those obtained for H$_2$O (including the increase of desorption efficiency throughout the first water ice electronic state), to the exception of the absolute photodesorption yield which is different.

The experimental results are summarized in table \ref{H2Oind_yields}, indicating the indirect desorption yields (as measured at the maximum of the first electronic state of water in the desorption spectrum, around 9 eV) for all investigated systems as well as the binding energy of each adsorbate to water ice and their mass. First we can observe that the indirect desorption yield goes in the order CO $>$ N$_2$ $>$ Ar $>$ Kr, and second that for a given adsorbate, the indirect desorption yield is systematically higher on D$_2$O ice than on H$_2$O ice. These facts, along with the observed linear increase of desorption efficiency through the first water ice electronic state in the desorption spectra, allow us to discuss the possible mechanisms of indirect desorption. 

Indirect desorption on water ice has been demonstrated before in the case of adsorbed benzene, using electron-stimulated desorption \cite{marchione2016,marchione2016a}. In that case the authors concluded that an electronic excitation transfer occurred between water and benzene. In our case, we can exclude any mechanism of that kind, i.e. that would involve energy transfer to the internal modes of the adsorbates. We observe indirect desorption occurring in the case of Ar and Kr below their excitation threshold: an electronic energy transfer is energetically impossible for these two adsorbates, and there is no notion of transfer to other modes (vibrational, librational) for these atoms as there could be for molecules. Let us now turn to the mechanisms that have been proposed for intact water desorption from the pure ice. Interestingly, these mechanisms predict different isotopic effects. The mechanisms of excited state repulsion and geminate recombination mentioned earlier predict no or close to no isotopic effect, while the H/D atom collision mechanism implies a higher efficiency of desorption for D$_2$O than for H$_2$O, because D atoms transfer momentum to heavier species more efficiently than H atoms. In the most recent experiments on pure water ice \cite{fillion2020}, we observed for intact water desorption a higher desorption yield for H$_2$O than for D$_2$O, in contradiction with several of these mechanisms. However the reverse isotopic effect is observed here: indirect desorption is more efficient on D$_2$O ices, implying a different dominant mechanism than for intact water desorption. The most fitting mechanism for indirect desorption is therefore energetic H/D atom collision, also called kick-out mechanism. 

\begin{table}
\caption{Summary of the investigated systems}
\def\arraystretch{1.2}
\label{H2Oind_yields}
\begin{tabular}{c c c c c}
\hline 
Adsorbate & \multicolumn{2}{c}{Indirect yield\footnote{H$_2$O-induced indirect desorption yield of the molecule for single layer systems at 9 eV.}} & Mass  & Binding energy\footnote{Binding energy on compact amorphous H$_2$O for 1 ML, from Smith et al. \cite{smith2016}.}  \\
 & \multicolumn{2}{c}{(mol/photon)} & (a.m.u.) & (meV)  \\
 & on H$_2$O & on D$_2$O & & \\
\hline

$^{13}$CO    & 3.4 $\times 10^{-3}$ & 4.5 $\times 10^{-3}$ & 29 & 122 \\
$^{15}$N$_2$ & 3 $\times 10^{-3}$   & 4 $\times 10^{-3}$   & 30 & 99  \\
Ar           & 1.4 $\times 10^{-3}$ & 3.8 $\times 10^{-3}$ & 40 & 75  \\
Kr           & 5 $\times 10^{-4}$   & 1.5 $\times 10^{-3}$ & 84 & 118 \\
\hline 
\end{tabular}
\end{table}

In the present case, where our adsorbates are atoms or diatomic molecules, this mechanism can be modelled as a pure kinetic energy transfer in a simple hard sphere collision. The fraction of transferred energy therefore simply depends on the masses of the colliding atom (H or D) and the adsorbate. It can be written as:

\begin{equation} \frac{E_{kin,transf}}{E_{kin,H/D}} = \frac{4m_{H/D}m_{ads}}{(m_{H/D}+m_{ads})^2} 
\label{sphd}
\end{equation}

Where E$_{kin,transf}$ is the kinetic energy transferred to the adsorbate upon collision, E$_{kin,H/D}$ the initial kinetic energy of the H/D atom, and m$_{H/D}$ and m$_{ads}$ the respective masses of the H/D atom and the adsorbate. Let us now estimate the kinetic energy of the H/D atom upon water dissociation. At a photon energy of 9 eV, considering a dissociation energy of 5.4 eV for H$_2$O and $\sim$ 0.7 eV going into OH translational and internal energy \cite{arasa2011}, the H atom ends up with approximately 2.9 eV of kinetic energy. A D atom would have slightly less. We can calculate that the absolute amount of kinetic energy transferred to the adsorbate therefore varies between 100 and 600 meV for our systems.

\begin{figure}
    \includegraphics[trim={0cm 0cm 0cm 0cm},clip,width=\linewidth]{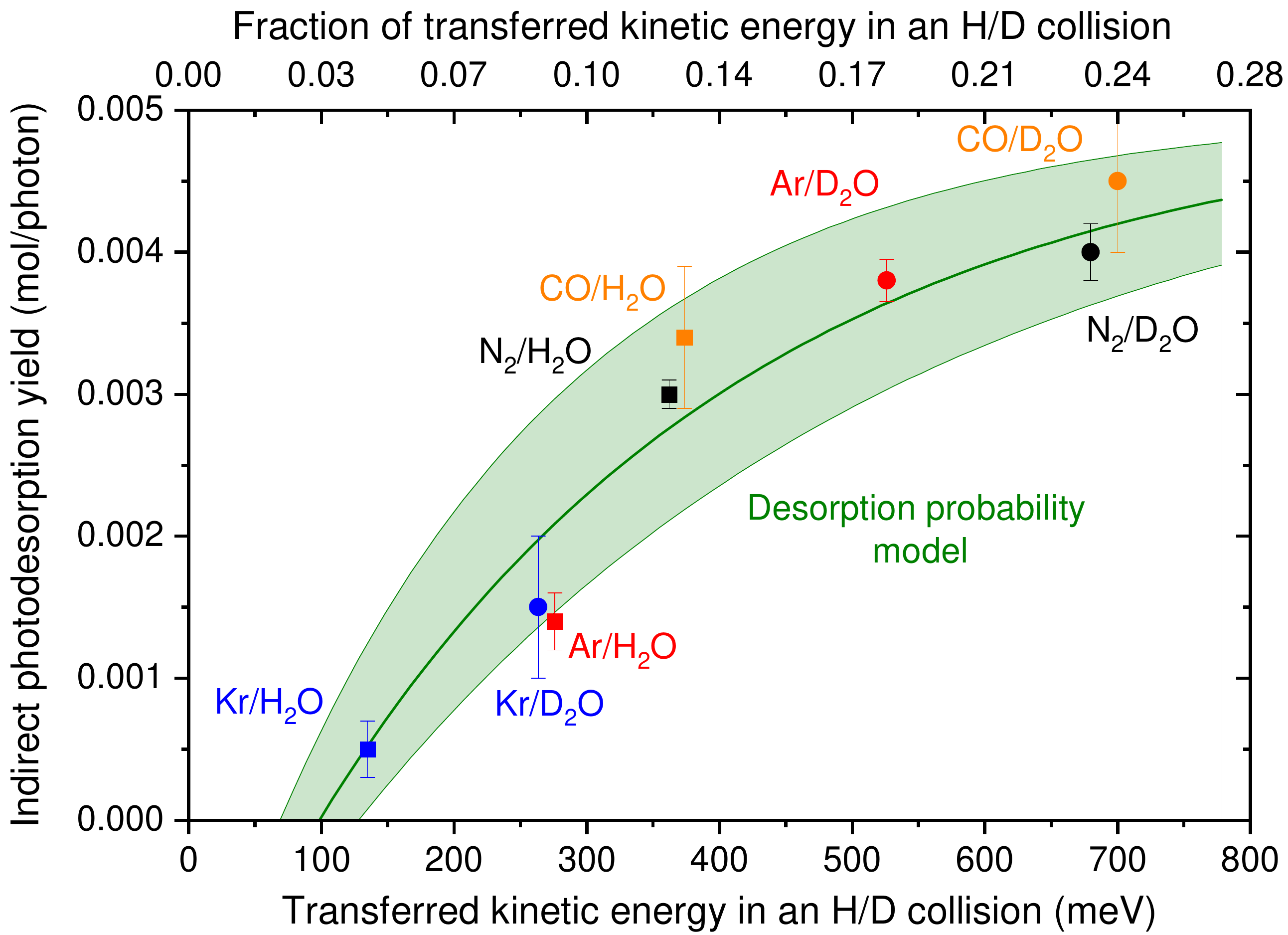}
    \caption{Indirect desorption yields for each system as a function of the fraction of kinetic energy transferred in the collision of an H/D atom with the adsorbate (see text for the formula). The indirect desorption yield is measured as the photodesorption yield at 9 eV at the maximum of the first electronic state of water ice. Error bars only reflect the signal to noise ratio of the desorption spectra, not errors in the relative calibration of the signals and coverage of the different species (see Supplemental Material). The green line is the desorption probability model from Fredon et al. \cite{fredon2017} for a binding energy value of E$_{bind}$ = 70, 100 or 130 meV (upper, middle and lower curves) and an initial H/D kinetic energy of 2.9 eV.}	
    \label{summary_des}
\end{figure}

Fig. \ref{summary_des} shows the indirect desorption yields of all investigated systems plotted against the estimated amount of kinetic energy that would be transferred to the adsorbate in a collision with an H/D atom (E$_{kin,transf}$). This figure shows a clear positive correlation. To confirm further the hypothesis of the collision mechanism, we need to know how the desorption probability of the adsorbate relates to the amount of transferred kinetic energy. This has been explored by Fredon et al. \cite{fredon2017,fredon2018} in molecular dynamics simulations. In these works the authors look at the desorption and energy dissipation of molecules at the surface of water ice which have been given translational energy. They derive an empirical function for the desorption probability as a function of initially imparted translational energy:

\begin{equation} P_{des}(E_{kin}) = 1 - exp\left(-\omega\frac{E_{kin}-E_{bind}}{E_{bind}}\right) 
\label{pdes}
\end{equation}

with P$_{des}$ the desorption probability, E$_{kin}$ the kinetic energy imparted initially to the adsorbate, E$_{bind}$ the binding energy of the adsorbate, and $\omega$ a factor set to 0.3 which fits best their simulation results. 
 
The desorption probability calculated from equation \ref{pdes}, is plotted on fig. \ref{summary_des} for E$_{bind}$ = 70, 100 and 130 meV and $\omega$ = 0.3. A good agreement with the experimental results is obtained, especially considering the error bars, relative calibration uncertainties and the fact that each species have slightly different binding energies. Some of the observed differences - e.g. the fact that the rare gases points tend to be closer to the lower curve and the diatomic points to the upper one, contrary to what would be expected from their binding energy differences - could indicate differences in the behaviour of atoms and molecules, as discussed below. The various sources of uncertainty however preclude firm conclusions.

The H/D collision mechanism also explains the linear increase of desorption efficiency observed between 8 and 9 eV (bottom panel of fig. \ref{Ar_H2O}). As photon energy increases throughout the first state, the amount of excess energy available increases as well and therefore the kinetic energy of the H/D atoms increases. This can (i) increase the amount of kinetic energy imparted by H/D atoms to the adsorbates, and (ii) increase the contribution of deeper layers of the ice to indirect desorption, as H/D atoms originating from deeper layers acquire sufficient kinetic energy to reach the surface and kick adsorbates. Both effects will lead to an increased desorption probability with photon energy. The changes in the desorption probabilities depend in principle on the different parameters in equation \ref{pdes} and should therefore vary for the different adsorbates, leading to different spectrum deformations. The desorption efficiencies are shown in the Supplemental Material. While the changes are indeed not exactly similar, they remain too small to analyse further.

Above 9 eV the increase in desorption efficiency is no longer observed, which can be explained by the loss of energy to other channels such as vibrational or electronic excitation of the fragments \citep{andersson2008}. Higher electronic states of water ice can be excited that also have a dissociative character \cite{schinke1995}, and therefore they can also lead to production of energetic H atoms and collision desorption.

One possibility that needs to be discussed before concluding definitively in favor of the H/D collision, is a mechanism that would involve collisions with OH/OD fragments. The more favourable mass ratio with the adsorbates means that even though OH/OD fragments acquire much less kinetic energy than H/D fragments upon dissociation, they would still transfer enough kinetic energy to induce desorption. The isotopic effect observed would then come not from the collision with the adsorbate, but from the photodissociation event, where OD fragments acquire almost twice more energy than OH fragment due to momentum conservation. Such a mechanism has however not been observed in the molecular dynamics simulations that observed the H/D collision mechanism. Furthermore, OH/OD fragments have a much lower mobility in ice \cite{arasa2010}, and from momentum conservation estimations \cite{arasa2011} would transfer twice less energy to adsorbates than in the equivalent H/D collision. We therefore consider this mechanism as possible but non-dominant. 

The full elucidation of the indirect desorption mechanism means that our results can be extrapolated to other systems. All weakly adsorbed atoms and diatomic molecules, at least, should follow a behaviour similar to what we observed. For these species the photodesorption yields can be derived from the desorption model of equations \ref{sphd} and \ref{pdes}, scaled to the experimentally measured absolute values. As pointed out above, there could be slight differences in the behaviour of atoms and diatomic molecules considering that for the latter excitation of internal degrees of freedom are possible. We would for sure expect deviations for even more sizeable molecules, for which a hard sphere collision model will not be appropriate any more, and for cases where a reactive collision can compete. Already for the case of H$_2$O desorption itself by the H collision mechanism, the theoretical calculations \cite{andersson2008} point to the importance of the collision happening at the center of mass of the molecule (on the O atom) for the efficiency of the process. For adsorbates that form hydrogen bonds with water we may expect also different processes coming into play, such as electronic energy transfer as in the case of benzene on water ice \cite{marchione2016}. Nonetheless, the present results could serve as a first basis for the modelling of indirect water-induced desorption in e.g. astrochemical models.

As we outlined previously, the so-called "kick-out" mechanism, which was among the first mechanisms suggested to explain desorption from pure water ice \cite{andersson2008}, had not received a completely unambiguous experimental demonstration of its existence so far, due to the complexity of disentangling the different possible mechanisms from the available data for pure water ice \cite{desimone2013,fillion2020}. We now provide such a demonstration in a slightly different system, weakly adsorbed molecules on water ice, for which all experimental facts converge. We thus shed light on a way indirect desorption can proceed. This is an important step towards a better comprehension of (indirect) desorption processes induced by electronic transitions and the fate of excitations at the water ice surface. This mechanism also deserves interest for any surface from which energetic H atoms could be produced (other hydride molecular ices, hydrogenated carbon surfaces...). 

\begin{acknowledgments}

We acknowledge SOLEIL for provision of synchrotron radiation facilities under the project 20180060 and we thank Laurent Nahon and the DESIRS beamline team for their help. This work was done in collaboration and with financial support by the European Organization for Nuclear Research (CERN) under the collaboration agreement KE3324/TE. The research was supported by the Programme National "Physique et Chimie du Milieu Interstellaire" (PCMI) of CNRS/INSU with INC/INP co-funded by CEA and CNES. Financial support from the LabEx MiChem, part of the French state funds managed by the ANR within the investissements d'avenir program under reference ANR-11-10EX-0004-02, and by the Ile-de-France region DIM ACAV program, is gratefully acknowledged. 

\end{acknowledgments}


\begin{thebibliography}{10}

\bibitem{vandishoeck2013}
E.~F. van Dishoeck, E.~Herbst, and D.~A. Neufeld,
\newblock Chemical Reviews {\bf 113}, 9043 (2013).

\bibitem{bartels-rausch2012}
T.~Bartels-Rausch {\em et~al.},
\newblock Reviews of Modern Physics {\bf 84}, 885 (2012).

\bibitem{hama2013}
T.~Hama and N.~Watanabe,
\newblock Chemical Reviews {\bf 113}, 8783 (2013).

\bibitem{kolb2010}
C.~E. Kolb {\em et~al.},
\newblock Atmos. Chem. Phys. {\bf 10}, 10561 (2010).

\bibitem{george2015}
C.~George, M.~Ammann, B.~D’Anna, D.~J. Donaldson, and S.~A. Nizkorodov,
\newblock Chem. Rev. {\bf 115}, 4218 (2015).

\bibitem{boogert2015}
A.~A. Boogert, P.~A. Gerakines, and D.~C. Whittet,
\newblock Annu. Rev. Astron. Astrophys. {\bf 53}, 541 (2015).

\bibitem{hogerheijde2011}
M.~R. Hogerheijde {\em et~al.},
\newblock Science {\bf 334}, 338 (2011).

\bibitem{caselli2012}
P.~Caselli {\em et~al.},
\newblock The Astrophysical Journal {\bf 759}, L37 (2012).

\bibitem{runne1995}
M.~Runne and G.~Zimmerer,
\newblock Nuclear Instruments and Methods in Physics Research Section B: Beam
  Interactions with Materials and Atoms {\bf 101}, 156 (1995).

\bibitem{reimann1988}
C.~T. Reimann, W.~L. Brown, and R.~E. Johnson,
\newblock Physical Review B {\bf 37}, 1455 (1988).

\bibitem{munozcaro2018a}
G.~M. Muñoz~Caro and R.~Martín~Doménech,
\newblock Photon-{Induced} {Desorption} {Processes} in {Astrophysical} {Ices},
\newblock in {\em Laboratory {Astrophysics}}, edited by G.~M. Muñoz~Caro and
  R.~Escribano Vol. 451, pp. 133--147, Springer International Publishing, Cham,
  2018.

\bibitem{mccoustra2017}
M.~McCoustra and J.~Thrower,
\newblock Exciton-{Promoted} {Desorption} {From} {Solid} {Water} {Surfaces},
\newblock in {\em Reference {Module} in {Chemistry}, {Molecular} {Sciences} and
  {Chemical} {Engineering}}, Elsevier, 2017.

\bibitem{oberg2016}
K.~I. Öberg,
\newblock Chemical Reviews {\bf 116}, 9631 (2016).

\bibitem{yabushita2013}
A.~Yabushita, T.~Hama, and M.~Kawasaki,
\newblock Journal of Photochemistry and Photobiology C: Photochemistry Reviews
  {\bf 16}, 46 (2013).

\bibitem{bertin2012}
M.~Bertin {\em et~al.},
\newblock Physical Chemistry Chemical Physics {\bf 14}, 9929 (2012).

\bibitem{bertin2013}
M.~Bertin {\em et~al.},
\newblock The Astrophysical Journal {\bf 779}, 120 (2013).

\bibitem{dupuy2017a}
R.~Dupuy {\em et~al.},
\newblock Astronomy \& Astrophysics {\bf 603}, A61 (2017).

\bibitem{feraud2019a}
G.~Féraud {\em et~al.},
\newblock ACS Earth Space Chem. {\bf 3}, 1135 (2019).

\bibitem{carrascosa2019}
H.~Carrascosa, L.~C. Hsiao, N.~E. Sie, G.~M. Muñoz Caro, and Y.~J. Chen,
\newblock Monthly Notices of the Royal Astronomical Society {\bf 486}, 1985
  (2019),
\newblock tex.ids: carrascosa2019 arXiv: 1903.11906.

\bibitem{johnson2011}
R.~E. Johnson,
\newblock Photolysis and radiolysis of water ice,
\newblock in {\em Physics and {Chemistry} at {Low} {Temperatures}}, pp.
  297--339, 2011.

\bibitem{ramaker1983a}
D.~E. Ramaker,
\newblock Chemical Physics {\bf 80}, 183 (1983).

\bibitem{dartois2015}
E.~Dartois {\em et~al.},
\newblock Astronomy \& Astrophysics {\bf 576}, A125 (2015).

\bibitem{marchione2016a}
D.~Marchione and M.~R.~S. McCoustra,
\newblock Phys. Chem. Chem. Phys. {\bf 18}, 29747 (2016).

\bibitem{yabushita2006}
A.~Yabushita, D.~Kanda, N.~Kawanaka, M.~Kawasaki, and M.~N.~R. Ashfold,
\newblock The Journal of Chemical Physics {\bf 125}, 133406 (2006).

\bibitem{miyazaki2020}
A.~Miyazaki {\em et~al.},
\newblock Phys. Rev. A {\bf 102}, 052822 (2020).

\bibitem{andersson2008}
S.~Andersson and E.~F. van Dishoeck,
\newblock Astronomy and Astrophysics {\bf 491}, 907 (2008).

\bibitem{arasa2015}
C.~Arasa, J.~Koning, G.-J. Kroes, C.~Walsh, and E.~F. van Dishoeck,
\newblock Astronomy \& Astrophysics {\bf 575}, A121 (2015).

\bibitem{yabushita2009}
A.~Yabushita {\em et~al.},
\newblock The Astrophysical Journal {\bf 699}, L80 (2009).

\bibitem{nishi1984}
N.~Nishi, H.~Shinohara, and T.~Okuyama,
\newblock The Journal of Chemical Physics {\bf 80}, 3898 (1984).

\bibitem{desimone2013}
A.~J. DeSimone, V.~D. Crowell, C.~D. Sherrill, and T.~M. Orlando,
\newblock The Journal of chemical physics {\bf 139}, 164702 (2013).

\bibitem{fillion2020}
J.-H. Fillion,
\newblock Astronomy \& Astrophysics  (2020),
\newblock Submitted, in revision.

\bibitem{westley1995}
M.~S. Westley, R.~A. Baragiola, R.~E. Johnson, and G.~A. Baratta,
\newblock Nature {\bf 373}, 405 (1995).

\bibitem{cruz-diaz2018}
G.~A. Cruz-Diaz, R.~Martín-Doménech, E.~Moreno, G.~M. Muñoz~Caro, and Y.-J.
  Chen,
\newblock Monthly Notices of the Royal Astronomical Society {\bf 474}, 3080
  (2018).

\bibitem{oberg2009c}
K.~I. Öberg, H.~Linnartz, R.~Visser, and E.~F. van Dishoeck,
\newblock The Astrophysical Journal {\bf 693}, 1209 (2009).

\bibitem{arasa2010}
C.~Arasa, S.~Andersson, H.~M. Cuppen, E.~F. van Dishoeck, and G.-J. Kroes,
\newblock The Journal of Chemical Physics {\bf 132}, 184510 (2010).

\bibitem{andersson2011}
S.~Andersson {\em et~al.},
\newblock Physical Chemistry Chemical Physics {\bf 13}, 15810 (2011).

\bibitem{avouris1989}
P.~Avouris and R.~E. Walkup,
\newblock Annual Review of Physical Chemistry {\bf 40}, 173 (1989).

\bibitem{nahon2012}
L.~Nahon {\em et~al.},
\newblock Journal of Synchrotron Radiation {\bf 19}, 508 (2012).

\bibitem{stevenson1999}
K.~P. Stevenson,
\newblock Science {\bf 283}, 1505 (1999).

\bibitem{doronin2015}
M.~Doronin, M.~Bertin, X.~Michaut, L.~Philippe, and J.-H. Fillion,
\newblock The Journal of Chemical Physics {\bf 143}, 084703 (2015).

\bibitem{dupuy2018c}
R.~Dupuy {\em et~al.},
\newblock Nature Astronomy {\bf 2}, 796 (2018).

\bibitem{doronin2015a}
M.~Doronin,
\newblock {\em Adsorption on interstellar analog surfaces: from atoms to
  organic molecules},
\newblock PhD thesis, Université Pierre et Marie Curie-Paris VI, 2015.

\bibitem{fayolle2011}
E.~C. Fayolle {\em et~al.},
\newblock The Astrophysical Journal {\bf 739}, L36 (2011).

\bibitem{orient1987}
O.~J. Orient and S.~K. Strivastava,
\newblock Journal of Physics B: Atomic and Molecular Physics {\bf 20}, 3923
  (1987).

\bibitem{rejoub2002}
R.~Rejoub, B.~G. Lindsay, and R.~F. Stebbings,
\newblock Physical Review A {\bf 65} (2002).

\bibitem{martin-domenech2015}
R.~Martín-Doménech {\em et~al.},
\newblock Astronomy \& Astrophysics {\bf 584}, A14 (2015).

\bibitem{cruz-diaz2014}
G.~A. Cruz-Diaz, G.~M.~M. Caro, and Y.-J. Chen,
\newblock Monthly Notices of the Royal Astronomical Society {\bf 439}, 2370
  (2014).

\bibitem{lu2008}
H.-C. Lu, H.-K. Chen, B.-M. Cheng, and J.~Ogilvie,
\newblock Spectrochimica Acta Part A: Molecular and Biomolecular Spectroscopy
  {\bf 71}, 1485 (2008).

\bibitem{marchione2016}
D.~Marchione, J.~D. Thrower, and M.~R.~S. McCoustra,
\newblock Phys. Chem. Chem. Phys. {\bf 18}, 4026 (2016).

\bibitem{smith2016}
R.~S. Smith, R.~A. May, and B.~D. Kay,
\newblock The Journal of Physical Chemistry B {\bf 120}, 1979 (2016).

\bibitem{arasa2011}
C.~Arasa, S.~Andersson, H.~M. Cuppen, E.~F. van Dishoeck, and G.~J. Kroes,
\newblock The Journal of Chemical Physics {\bf 134}, 164503 (2011).

\bibitem{fredon2017}
A.~Fredon, T.~Lamberts, and H.~M. Cuppen,
\newblock ApJ {\bf 849}, 125 (2017).

\bibitem{fredon2018}
A.~Fredon and H.~M. Cuppen,
\newblock Physical Chemistry Chemical Physics {\bf 20}, 5569 (2018).

\bibitem{schinke1995}
R.~Schinke,
\newblock {\em Photodissociation dynamics: spectroscopy and fragmentation of
  small polyatomic molecules}, Cambridge monographs on atomic, molecular and
  chemical physics No. ~1 (Cambridge University Press, Cambridge, 1995).

\end{thebibliography}
\end{document}